# Evolutionary algorithm based adaptive navigation in information retrieval interfaces

Dmytro Filatov and Taras Filatov PhD, Ukraine

**Abstract**
In computer interfaces in general, especially in information retrieval tasks, it is important to be able to quickly find and retrieve information. State of the art approach, used, for example, in search engines, is not effective as it introduces losses of meanings due to context to keywords back and forth translation. Authors argue it increases the time and reduces the accuracy of information retrieval compared to what it could be in the system that employs modern information retrieval and text mining methods while presenting results in an adaptive human-computer interface where system effectively learns what operator needs through iterative interaction.  In current work, a combination of adaptive navigational interface and real time collaborative feedback analysis for documents relevance weighting is proposed as an viable alternative to prevailing 'telegraphic' approach in information retrieval systems. Adaptive navigation is provided through a dynamic links panel controlled by an evolutionary algorithm. Documents relevance is initially established with standard information retrieval techniques and is further refined in real time through interaction of users with the system. Introduced concepts of multidimensional Knowledge Map and Weighted Point of Interest allow finding relevant documents and users with common interests through a trivial calculation. Browsing search approach, the ability of the algorithm to adapt navigation to users interests, collaborative refinement and the self-organising features of the system are the main factors making such architecture effective in various fields where non-structured knowledge shall be represented to the users.
Keywords: evolutionary, adaptive, search engines, navigation, web, collaborative, social, interface, information retrieval, knowledge, document, algorithm, adaptive, refinement, text mining.

**1 Introduction**

In the world where indexing approach and linguistic search prevail the habits of search and location of required information are well established. According to Meadow [19], there are four different types of search, which are:
1) known-item search (looking for a specific source where desired information is known to be stored)
2) specific-information search (looking for a specific information, the goal of the search can be formulated)
3) general-information search (goal exists but is hard to formulate, "I know when I see it")
4) searching to explore the database (goal not specified, users wants to get familiar with the knowledge corpus )

From the point of view of implementation in knowledge acquisition and representation systems, there are basically two known approaches of how users are allowed to perform their searches:
a) indexing, employed in the state-of-the-art search engines, where users have to type in key phrases i.e. they have to formulate their request linguistically, in a 'telegraphic' way;
b) browsing, employed in end web sites, knowledge bases, documents collections, directories, where users have to navigate through massive hierarchy in order to find the information they want;

The first approach is being used for indexing of the whole internet in the state-of-the-art search engines. The main advantage of this approach is full automation as it is designed for indexing huge amounts of text documents available in the internet. The disadvantages are those that are caused by automation and working on a global scale. Firstly, the problems of text understanding and natural language processing are one of the most challenging in Artificial Intelligence field and still remain without an efficient solution. Adjacent are the problems of classification and relevance calculation, the so-called 'web clustering' problem [2]. Secondly, there is a limit in complexity of the algorithms. The latest Information Retrieval (IR) methods cannot be applied on a full scale as there always should be a compromise between speed and accuracy of indexing. Despite all the disadvantages, the indexing approach prevails and serves especially well for the known-item and specific-information search types.

The second approach is being used where manual indexing is possible. The advantage is that users get ready-to-use taxonomy i.e. structured navigation for the documents corpus. Such index is usually more accurate and understandable for users. The disadvantages are that the amount of human labour required to create and maintain such index grows in a progression with the growth of the corpus. The possibility for human errors is also present. The main disadvantage for users is usually slow updating. It is more likely that automated crawler will spot new or updated documents faster than human operator. For the explorative tasks ('general-information search', 'searching to explore the database'), however, the browsing approach of organizing the navigation is by no means the best solution. Obviously, when the user is driven by the third type of search ("I know it when I see it"), numerous factors, coming from the organization of navigation and semantic links in the particular knowledge representation system, play role varying the chances for the user to get the desired information.

It can be concluded therefore that the general-information search is presently the one which is less supported by knowledge acquisition and representation systems while it still remains to be one of the natural ways of searching the information.

It is logical to suppose that there could be hybrid solutions employing the advantages of different approaches in order to serve the needs of such users better. E.g., browsing approach to navigation could be combined with automated indexing. It would be possible to use the latest IR and Data Mining approaches in order to refine results and propose most relevant documents using the adaptation to current user's needs and other users' feedback.

Despite the demand in such applications seems to be high, there are no well known hybrid solutions to be applied widely. We believe this is due to the fact that such systems are of a more complex nature and there was no right solution yet proposed.

The idea of automated and user-oriented data pre-processing solutions for web is not novel. A number of systems were developed to organise the information requested by users into semantic data collections [1, 15, 17, 24, 27] which makes it more convenient to browse and search for related documents. In these cases, however, users need to know (or predict) the wording that is used in the documents they are looking for in order to specify a search query to be passed to the search engine. On the other hand, in some systems users are faced with knowledge base with a host of categories, which may make the search confusing. We believe it is better to propose a random set of documents initially; this set to be of proportional size to the navigational panels most web surfers are used to; then make the system gradually recognize users' interests and modify the set correspondingly. Resembling approach was presented in [28] where a method for dynamic link generation is proposed. The online module, however, was not developed. No means for automated data collection have been though of. In addition, there is no mechanism mentioned to remember user so each time the user is new to the system which reduces the possibilities for analysis and adaptation. The mechanism for links fetching is not presented in detail, and no ultimate algorithm is proposed.

The main obstacle, we may assume, for the further development and implementation of such systems is the involvement of technologies from various fields. It is necessary to provide a solution for automated data collection, data storage, compression and retrieval, user identification, user-system and user-user interaction, information retrieval, documents analysis, relevance calculation, adaptation, access patterns analysis etc. That is why the successful researches in this field are limited.

There are, however, few researches that have achieved to present systems of such scale. As a typical representative we will briefly overview the multi-agent collaborative Web mining system [9]. Authors position their system as a tool for web content mining and post-retrieval analysis. System employs multi agent structure and provides strong collaborative functionality allowing users to re-use the results of searches performed by other users. The shortcomings of the system, from our point of view, are the following. First, special software is required to use the system. It would be logical to suppose that it is more convenient for the user to use a standard web browser and surf the internet in a standard way while the assistant system should be represented by an interface occupying just small part of the screen. Second go related issues of users' identification and calculation of relevance between users and documents. Users, which start to use the system, are required to register and specify their topic of interest. The web 2.0 and social web trends of nowadays which promote usability and ease of use make internet surfers choose systems which are able to identify them automatically and adapt to their needs during the interaction. We believe a collaborative information retrieval system could work with a minimal user input while taking maximum from users trivial browsing interactions. Regarding the relevance calculation, in the system proposed by Chau et al [9] there is no mathematical model to calculate the relevance between users and documents. There is a concept of 'Knowledge dashboard' where the links found by the users researching the same area appear. What is going to happen when user has multiple areas of interest? What if users change their mind and get interested in other topic? What if there are more related links than there could be displayed at one screen? How to weight the similarity of interests between users, the relevance of pages and the actuality of the information for the user? The multi-agent collaborative Web mining system is a significant achievement however it does not give answers to these questions. The issue of data storage goes adjacent with the data processing and calculation. We believe a method can be found to map the documents of the corpus into a single multidimensional space which will enable trivial mathematical calculations to be used to resolve some of the abovementioned questions.

**2 Main principles**

These are the main principles we have followed when creating our system:

1. Data preparation and relevance calculation. Unstructured data from given sources may be refined using the Information Retrieval techniques in order to provide the system with the preliminary understanding of the knowledge corpus. State of the art techniques may also be used to calculate the relevance of documents in real time mode.

2. Adaptive interaction. We propose a system for interaction based on evolutionary algorithm powered navigational panel with features of adaptation to current user behaviour and needs.

3. Further refinement through users feedback. In the system proposed certain weights and features as well as additional interfaces correspond to social factor making use of users interaction in order to improve the service. i.e. users help the system to understand better the taxonomy of the knowledge it possesses.

4. Collaborative factor. The system proposed learns through the history of interaction the interests of users and is then able:

a) to propose relevant documents based on previous interests indicated by the user;
b) to match users by interests and suggest documents the user might be interested in, by using the research carried out by other users with similar interests.

**3 The system**

**3.1 Data acquisition**

The data acquisition does not lie within a focus of the current paper and therefore we will mention it briefly here. For our experimental implementation we have created a typical search engine-like system for the purpose of data collection and preprocessing. The system consists of a data collection and processing module that crawls and indexes the documents available via HTTP protocol (i.e. web pages) starting with the given start pages and limiting the crawler to stay within the given domains. The documents are then processed through Porter's stemming algorithm [23] and the stop words filter and are reproduced as vectors using the tf.idf metrics. [25, 26] To minimize the calculation time the vector space is reduced using the linear dimensionality reduction techniques such as Principal Component Analysis (PCA) which are known to give better results for such data [18]. The process of calculating the intrinsic dimensionality of the corpus precedes the actual dimensionality reduction stage which improves the efficiency of the mapping. The relevance between the documents established using the *Knowledge Map* compressed to the intrinsic dimensionality of the corpus has sometimes turned out to be closer to the real, actually known, relevance, than the same calculated using the uncompressed corpus during our experiments (more details in chapter 4 below). A detailed description of data acquisition and storage used in our system starting from searching and indexing techniques to comparison of various dimensionality reduction methods would be out of scope of the current paper as here we focus on the innovative parts of evolutionary navigation so only the key principles of data acquisition part are explained. It is important to add however that PCA has been chosen as it returned the most stable variance of results when evaluated through relevance comparison with the known data. Also important is the fact the PCA itself enables calculation of the intrinsic dimensionality [10]. The procession of mapping or scaling is the final stage in the data acquisition part of processes in our system. The obtained multidimensional vector space is used as a *Knowledge Map* for the corpus of indexed documents and the Euclidian distance between the vectors of corresponding documents is used as a metric for relevance. Further justification of the method is given in chapter 4 where the results of experiment are provided.

### 3.2 Interface

In pursuit of the task of creating a hybrid solution serving the needs of users motivated by 'general information search' we have learned that one of the most important initial tasks was to create a proper interface layout. After the study of research works devoted to human-computer interaction and usability, few key principles have been outlined which have formed the interface of our experimental model. It is known [12, 13, 21] that navigation area should be a compact block, which is possible to overview with a single eye movement. Human brain is able to keep track of a limited number of objects, according to studies [3]. The navigational panel should not change its location and remain at the same place, closer to the top of the page [4, 6, 7, 13]. Placing the panel in the left side of the screen results in better performance and improved navigation times [5, 14, 16]. Following these principles we have designed the layout for the interface of our system.

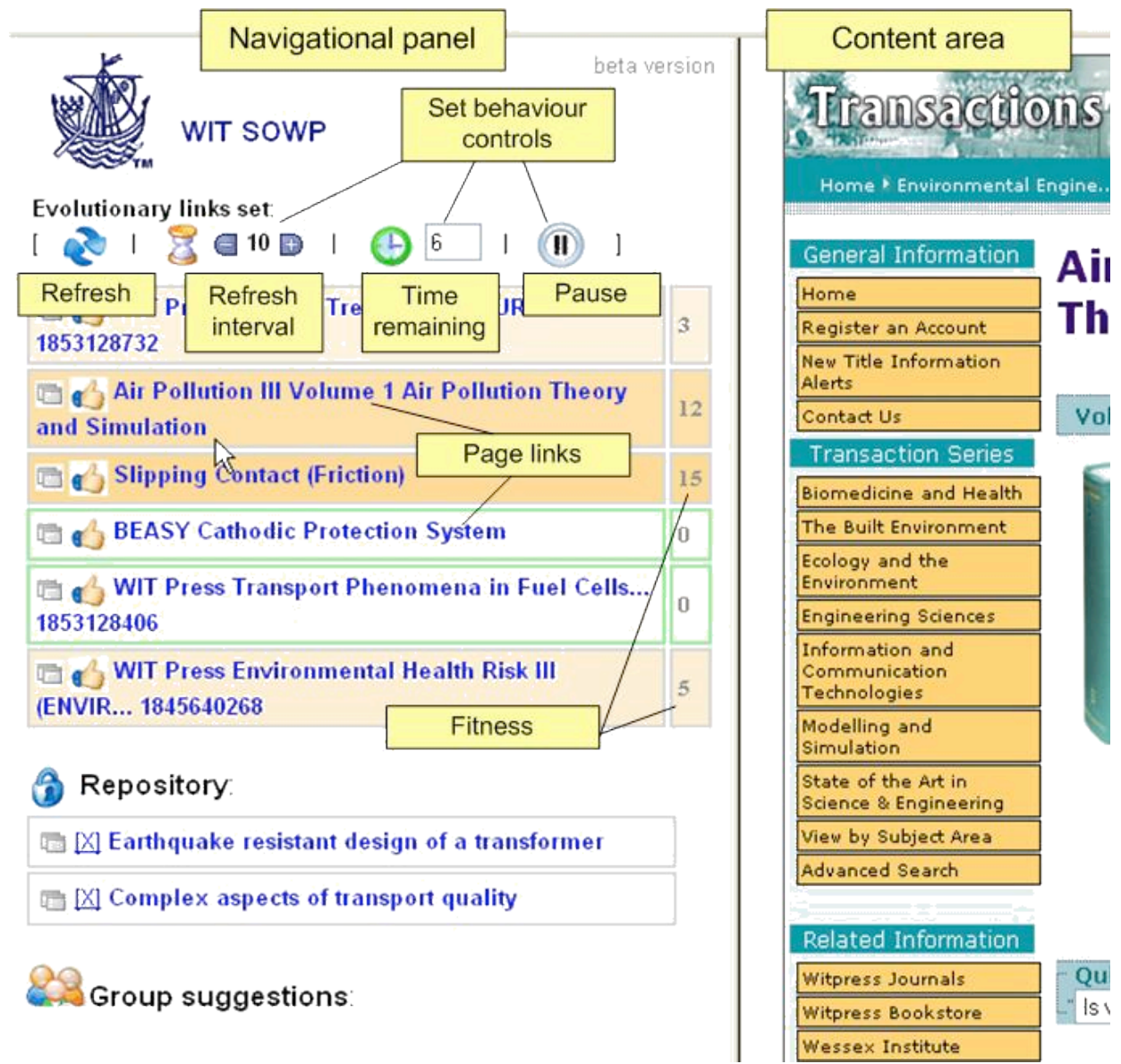

Figure 3.1 Navigational panel interface. Implementation.

The content area of the web browser has been divided in two vertical column frames. The first, leftmost frame, is of fixed width and contains the navigational panel. The remaining space is a content area. Navigational panel remains at its place all the time. It is powered by our system. The content area displays different web pages chosen by user with the help of the navigation panel. The interface of the navigational panel consists of the following elements:

**3.2.1 Set behaviour controls**

'Refresh' button will stop the process of evolution in the Set and fill it with random, zero fitness pages.
'Refresh interval' controls allow to adjust the time between iterations of the process of evolution of the Set.
'Time remaining' control allows user to see how much time in seconds is left before the Set will update itself and the following iteration of the evolution will occur.
'Pause' button allows to 'freeze' the process of evolution so that the Set remains in the same condition if the user wants so.

The Set itself consists of a fixed number of elements representing the pages indexed by the system. Each element consists of:

### 3.2.2 Set elements

'Open in a new window' button which opens the page in a new window.
'Add to Favorites' button saves the page to Favorites section.
'Page link' is a title of the page linked to the page itself. When clicked, it opens the page contents in the content area.
'Fitness' column displays the value of the current page's fitness calculated by the evolutionary algorithm powering the Set.

### 3.2.2. Favorites section

'Favorites' is a section where pages are saved by the user. While the Set contents are being changed over time according to algorithm, the 'Favorites' section remains stable so that user is able to save interesting pages there.

### 3.2.3. Social suggestions section

'Social suggestions' is a section which is, again, controlled by the algorithm. The system calculates which of the other users have similar interests and displays the pages from their 'Favorites' in this section.

## 3.3 Interactive part

### 3.3.1. The Set

Following our interface specification, we have a navigational panel of a limited size (7 to 10 items in our experiments, in accordance to [3]). Our aim is to use this limited area $Set_{size}$ in a best way to provide users with relevant results.

### 3.3.2. Initial stage

In case the user hasn't visited the system before and hasn't made any action in the system yet there is simply no data for the system to analyze user's interests.
Therefore, following our browsing approach, the system has to give the user something to start with. As much topics as possible should be proposed within the given limited items number and due to this limitation it would be logical to present the most unrelated links.
Let's divide the *Knowledge Map* into $Set_{size}$ number of clusters and pop out a random link from each of the clusters. Let's introduce the *random_links_show* function to be used when system knows nothing about current user:

  random_links_show()  (3.3.2.1)
$$ClusterN_{size} = \frac{KnowledgeMap_{size}}{Set_{size}}$$
    for N = 1 to $Set_{size}$
       {
         $linkN = random(ClusterN)$
       }

In case there is no response from the user after some time R (response time), it is likely that user is not interested in the topics proposed, in which case the process is repeated and new random links are proposed. It will however look more natural and psychologically easier to perceive for the user if we replace a small number of links and leave others. [8, 11]

$Links_{replace}$ = const (3.3.2.2)

$Links_{replace} << Set_{size}$

In our experimental implementation a default value for $Links_{replace}$ has been set to 3.

Thus at each iteration we will replace *links_replace* number of random links from the set with new links according to random_link_show() function.

The algorithm for set renewal will therefore look as following:

*random_links_show()* (3.3.2.3)

$$ClusterN_{size} = \frac{KnowledgeMap_{size}}{Set_{size}}$$

$LR_{array}$ = { select $Links_{replace}$ number of random id's from the Set } for N = 1 to $Set_{size}$
{
  if $linkN \notin LR_{array}$
  {
    $linkN = random(ClusterN)$
  }
}

This way our algorithm becomes evolutionary and *links_replace* provides a mechanism for recombination. The algorithm renews itself in time trying to display the most scarcely distributed topics from the corpus and in such way tries to invite user for interaction.

### 3.3.3 Interaction

When user is interested in a certain topic, he/she will choose the corresponding link. The system has to use this important information. After first interaction of such kind has occurred, the stage of adaptation and supervised learning begins. The system has to adapt to current users needs and also learn better the characteristics of current user and also the characteristics of knowledge corpus (using the feedback received as interaction from current user).We will now introduce an *interactive_iteration*() function which in contrast to *random_links_show* will take into account the user's feedback, current Set contents and previously collected data.

### 3.3.4 Fitness

We need somehow to distinguish the links the user is interested in from links for which user has shown no interest. Let's introduce a traditional parameter of evolutionary algorithms, the fitness, into our system. Let's make it that when the user clicks a link, the fitness of the corresponding link will increase at $fitness_{click\_modifier}$ value.

$$fitness_{click\_modifier}\ const \qquad (3.3.2.4)$$

In our experiments the $fitness_{click\_modifier}$ has been equal to $Set_{size}$.

Let's also state that links with positive fitness remain in the Set and the recombination rule (links_replace in *random_links_show()* function) applies only to those links with minimal or zero fitness, e.g. now the less fit, not random, links will be replaced. This will give users time to study properly the selected links and will also allow further adaptation and complex evolution schemes.

### 3.3.5 Ageing (fitness penalization)

To emulate the natural processes of distraction and vaporization of interest and also to anticipate the moment when the user wants to switch to other topic we will penalize the positive fitness with time. Let's therefore during each minimal time interval (*ageing_interval*) decrease the fitness of all links with positive fitness by 1.

ageing_interval const (3.3.2.5)
ageing_interval << refresh_interval

```
ageing()
    if (ageing_interval)
    {
      for N = 1 to Set_size
        {
          linkN_fitness = linkN_fitness - 1
        }
    }
```

To adapt to user's needs let's also compliment our algorithm with another function in addition to *random_links_show*. Would be logical after a certain time of interaction for a system to adapt and make it so that new emerging links are not random but most relevant to the topic the user is currently interested in. We can assume that links with positive fitness represent current user's interest with a certain degree of accuracy. We describe the improved algorithm after we introduce a mechanism of Favorites below.

### 3.3.6 Favorites

We add a mechanism of Favorites for two reasons:
a) Users convenience so that they may store links of current interest in a separate place where links are not subject to ageing and will therefore not disappear until user decides to remove them manually.
b) 'Favorites' mechanism is also very important information used by system for learning and adaptation as it represents a confirmed expression of user's interest.

Let's allow Favorites influent the evolution process as if they were present in the links set and had comparatively high ranks

$$Favorites_{fitness} = \max ( Favorites_{fitness\_const}, 2* \max( Link1_{fitness}, ..., LinkN_{fitness} )) \quad (3.3.2.6)$$

In our experimental implementation we set $Favorites_{fitness\_const}$ to 50.

This will allow users control the process of evolution better by adding and removing links from Favorites.

### 3.3.7 A Weighted Point of Interest

As there could be more than one link with positive fit in the Set, here opens a unique possibility for a user to express their interest in different, even not related topics within the corpus. Our system should make use of it and try to find documents simultaneously related to all the topics of interest.

Operating with multidimensional space of our *Knowledge Map* we may assume that point laying at equal distance from the points of all of the links with positive fitness will represent the centre of interest.

In order to consider the different level of interest of different links let's move the coordinates of this point of interest closer to those points corresponding to links with higher fitness. We therefore will establish the coordinates for *Weighted Point of Interest* (*WPI*).

The function to calculate the *WPI* for the current user is given below.

*Calculate_WPI(current user)* $\hspace{5cm}$ (3.3.2.7)

for D = 1 to $KnowledgeMap_{dimensionality}$
  for N = 1 to $Set_{size}$
    $WPIcoord_D = WPIcoord_D + LinkN_{coord[D]} * LinkN_{fitness}$
  for F = 1 to $Favorites_{size}$
    $WPIcoord_D = WPIcoord_D + LinkF_{coord[D]} * LinkF_{fitness}$

$$WPIcoord_D = \frac{WPIcoord_D}{Set_{size} + Favorites_{size} + \sum_{M=1..N} LinkM_{fitness} + \sum_{G=1..F} LinkG_{fitness}}$$

Having established the coordinates of WPI we will assume this is a central point of interest of the user and the most relevant documents will have a minimal Euclidian distance to that point. Thus, if there are links with positive fitness in the set or there are links in Favorites we replace the random function with the function of relevance:

```
  if positive fitness                                         (3.3.2.8)
    link[N] = most_relevant (WPI)
  else
    link[N] = random (cluster[N])
```

Where most_relevant() function basically represents choosing a document in the collection with the closest Euclidian coordinates to those of the WPI.

### 3.3.8 Mutation

It is difficult for the system to determine when user changes her mind and is not interested in the topic anymore. Two mechanisms help to deal with such situation:
 a) Ageing - with time the fitness of all links in the set will come to zero and the system will return to the initial stage with random links.
 b) Reset control – special control in the navigational panel which allows user to reset (fill the set with random, zero ranked links) to the initial state at any time.

These mechanisms, however, won't help to construct complex search requests as once user expresses her interest in one topic, the system gets oriented at that topic only. The new emerging links will be related to that single topic and the possibility to research related and unrelated topics simultaneously in order to make use of WPI mechanism becomes obscured.

That makes reasonable for us to introduce another traditional element of evolutionary algorithms which is mutation.

We adjust the algorithm so that even when there are positive fitness links in the Set and there are links in the Favorites repository, still there is a probability for the random link to appear.

Let's introduce the parameter which will determine with what probability the emerging links will undergo mutation.

$$Mutation_{probability} = \text{const} \tag{3.3.2.9}$$

In our experimental implementation we set $Mutation_{probability}$ to 0.3 which has been approved by the experiments to be an optimal value.

*interactive_iteration*() (3.3.2.10)

$$ClusterN_{size} = \frac{KnowledgeMap_{size}}{Set_{size}}$$

for N = 1 to set_size
    if ( random(0..1) $\leq Mutation_{probability}$ ) then $linkN = random(ClusterN)$

### 3.3.9 User adaptation

User identification and authentication mechanism is not a subject of this paper and may be realized using traditional mechanisms widely applied in client server applications. We have initially realized the user identification in our experimental system through login-password authentication combined with the mechanism of sessions. This authentication mechanism has been later replaced with the mechanism of cookies. The latter approach is better for research and evaluation purposes as users do not have to log in, which simplifies the work with the system. In applied implementations the authentication would be most likely handled by externals mechanisms as such system as described in this work would be integrated with existing user bases e.g. Facebook Social Graph.

In order to propose a broader variety of links to choose for the user it would be better not to display the links which have been already shown recently. We can see here it is necessary for the system to remember the history of what it had shown to the user and when:

history_links(user_id, link_id, timestamp)                                    (3.3.2.11)

We can also introduce the *history_recent* parameter trying to estimate after how much iteration the link may become actual for the user again.

$History\_recent_{iterations}$ = const                                          (3.3.2.12)

In our experimental implementation we set $History\_recent_{iterations}$ to 20.
Let's add the following condition to our Iteration Function then:

if history_displayed(timestamp) ≠ [NOW - history_recent(iterations_number)]    (3.3.2.13)

### 3.3.10 Favorites

We may also record for how long and which links have been present in the Favorites. These links are confirmed to be interesting for the user. The more they stay in the Favorites the more interesting and relevant they tend to be for the user. Our system should therefore record this data and use it for Social suggestions mechanism to share the experience of the current user with other users of the system.

Therefore, let's introduce a parameter *time_alive*, reflecting user's interest in particular link expressed by the time it has been stored in the Favorites:

favorites_history(user_id, link_id, time_alive, timestamp)

*Time_alive* will increment while link stays in the Favorites.
*Timestamp* will store the system's date and time when the record is updated.

### 3.3.11 Dormant mode

As user may not always pay attention to the system while it is evolving and history is still being recorded, this may cause the system to collect wrong data presuming user is still proactive. We propose to introduce a concept of Dormant mode. This mode will activate when user makes no actions in the system during *dormant_count* seconds.

It would be logical to vary the dormant count depending on the size of the document the user is currently reading and also on individual user's characteristics that reflect his/her level of activity within the system. Such calculations will increase the complexity of the system therefore in current implementation we propose a constant value for *dormant_count*:

dormant_count = const                                                          (3.3.2.14)

In our experimental implementation the *dormant_count* parameter has been set to 300.
In turn, this will stop the incrementing of *time_alive* which should also be reflected in our algorithm. Let's also penalize the *time_alive* gained by links staying in Favorites during the time of inactivity:

if inactive time > dormant_count                                               (3.3.2.15)
  then
    dormant_mode = on
    for all link_id in Favorites
      users_interests(interest_level) = users_interests(interest_level) - dormant_count

### 3.3.12 Social suggestions

The system stores the history of using the Favorites mechanism for each user. This data may be mined to extract useful knowledge about the interests of particular users and interchange information between users that seem to have similar interests. This will enable indirect collaboration between the users of the system as new users may follow the successful searches performed previously by other users and therefore get to study the required topic in a faster and deeper way. [22] There are two directions here for the refinement of results using data mining approach:
1. To understand the current topic of interest of the user.
2. To find users with similar interests and fetch the links they have found useful.

In our system the user may indicate his/her interest in the link by clicking its title in the Set. After such an action the link will gain the positive rank and will remain in the Set so that the user has time to study the document. This will also attract other related documents into the Set. The user may however decide after having examined the document that the link is irrelevant to their search. Thus we cannot use the history of interim interactions in the Set as indication of users interests. The mechanism of Favorites is a more trustworthy source of such information, as the links in the Favorites are added manually by users and therefore are confirmed to be of interest.

To understand the current topic of interest of the user we may use two sources:
   a) the actual data i.e. links present in the Favorites now
   b) the history record of Favorites

In order to give user more control over Social suggestions mechanism, let's implement two modes determining the source to be used in order to establish the topic of interest:
   a) when there are no links in the Favorites, the interests of the current user are determined using the history of the Favorites (if the number of records is > 0)
   b) when there are links in the Favorites, the history is not taken into account and the Social suggestions are based on what is present in the Favorites at the current moment.

The following data format is being logged:

favorites_history(user_id, link_id, time_alive, timestamp)          (3.3.2.16)

The important parameters are *time_alive* and *timestamp*. *Time_alive* gives us information on how the user has estimated the importance of the link. Different users may have different activity level, the speed of reading, searching/browsing habits etc. Let's normalize *time_alive* parameter for each link by setting a maximum *time_alive* of any link stored by current user during all the time to 1 and finding a proportional value for each link:

$$LinkF_{timealive\ modifier} = \frac{LinkF_{timealive}}{\max_{N=1..F} LinkN_{timealive}}$$          (3.3.2.17)

Where F is an identifier of the link stored in Favorites history of the current user.

Timestamp allows us to penalize the old history following the assumption that old interests are less actual. We introduce a modifier value (a multiplier of link importance), which will be near to 1 for newest links and near to 0 for oldest links.

$$LinkF_{age\_modifier} = \frac{LinkF_{timestamp}}{NOW() - T_0} \qquad (3.3.2.18)$$

where $T_0$ is the timestamp of the oldest link (first link added to Favorites by the current user), NOW() is the timestamp of the current moment and $LinkF_{timestamp}$ is the timestamp of the moment when the history record of the link *F* has been updated.

Using either a) or b) approach let's establish the Weighted Point of Interest (WPI) using either links actually present in Favorites or normalized history data:

```
if F > 0                                                                (3.3.2.19)
{
    For D = 1 .. KnowledgeMap_dimensionality
    {
```
$$WPI(social)_{coord[D]} = \sum_{N=1..Favorites_{size}} \frac{linkN_{coord[D]}}{N}$$
```
    }
}
else
{
    For D = 1 .. KnowledgeMap_dimensionality
    {
```
$$WPI(social)_{coord[D]} = \frac{\sum_{F=1..History_{size}} LinkF_{timealive\_modifier} * LinkF_{age\_modifier} * LinkF_{coordD}}{\sum_{F=1..History_{size}} LinkF_{timealive\_modifier} * LinkF_{age\_modifier}}$$
```
    }
}
```

We need now to find users who match current user with their interests. It is obvious that in order to make such calculation, the system should store the actual WPIs i.e. the normalized coordinates of interest for each user. The coefficients for normalization are the following:
- Time (1 = most recent, 0 = the oldest record)
- Time spent in the Favorites (1 = the longest, 0 = the shortest). Before comparison the values for each user should be normalized so that 1 is the link which has spent the longest time in the Favorites of the certain user and 0 is the link with the shortest time spent in
  the Favorites of the certain user.

The WPI coordinates normalized by the abovementioned parameters ideally represent the current interest of the user. These values, recalculated periodically, should be stored in a separate database record for each user to enable fast real time calculation for the social suggestions and other algorithms.

Back to our social suggestions algorithm, let's calculate the level of interest of each page for the current user.

For N = 1 to $Pages_{total}$ (N $\notin$ Favorites) \qquad (3.3.2.20)

```
{
  For U = 1 to Users_total
  {
    PageN_LOI = PageN_LOI + LOI_modifier_N,U
  }
}
```

Where $Users_{total}$ is the number of users in the database.

$Pages_{total}$ is the number of pages in the database.

$LOI\_modifier_{N,U}$ is a coefficient reflecting how the level of interest of the certain document N for the certain user U should affect the level of interest of the same document for the current user. This value is a multiply of the following coefficients:
- Time (1 = most recent, 0 = the oldest record)
- Time spent in the Favorites of the user U (1 = the longest, 0 = the shortest), normalized
- Distance. Euclidian distance between the WPI of the current user and the WPI of user U.

$$LOI\_modifier_{N,U} = Time_{N,U} * Time\_alive_{N,U} * Distance(WPI_{current\_user}, WPI_U) \quad (3.3.2.21)$$

The simple algorithm listed above will summarize the level of interest for each page as it should be for the current user taking into account both the active feedback of the other users (pages placed in Favorites) and passive feedback (time spent by the pages in the Favorites excluding 'dormant' periods, number of users suggesting the same page, the percentage of matching interests between the current user and the suggesting users, the overall levels of activity of the suggesting users, comparative novice of the information etc).

### 3.4 Final algorithm

Let us now conclude with a final algorithm for the functionality of our system. We propose two flowchart illustrations for the algorithm of the described system. The first flowchart represents periodical process maintaining the system which is initialized every minimal period of time (1 second):

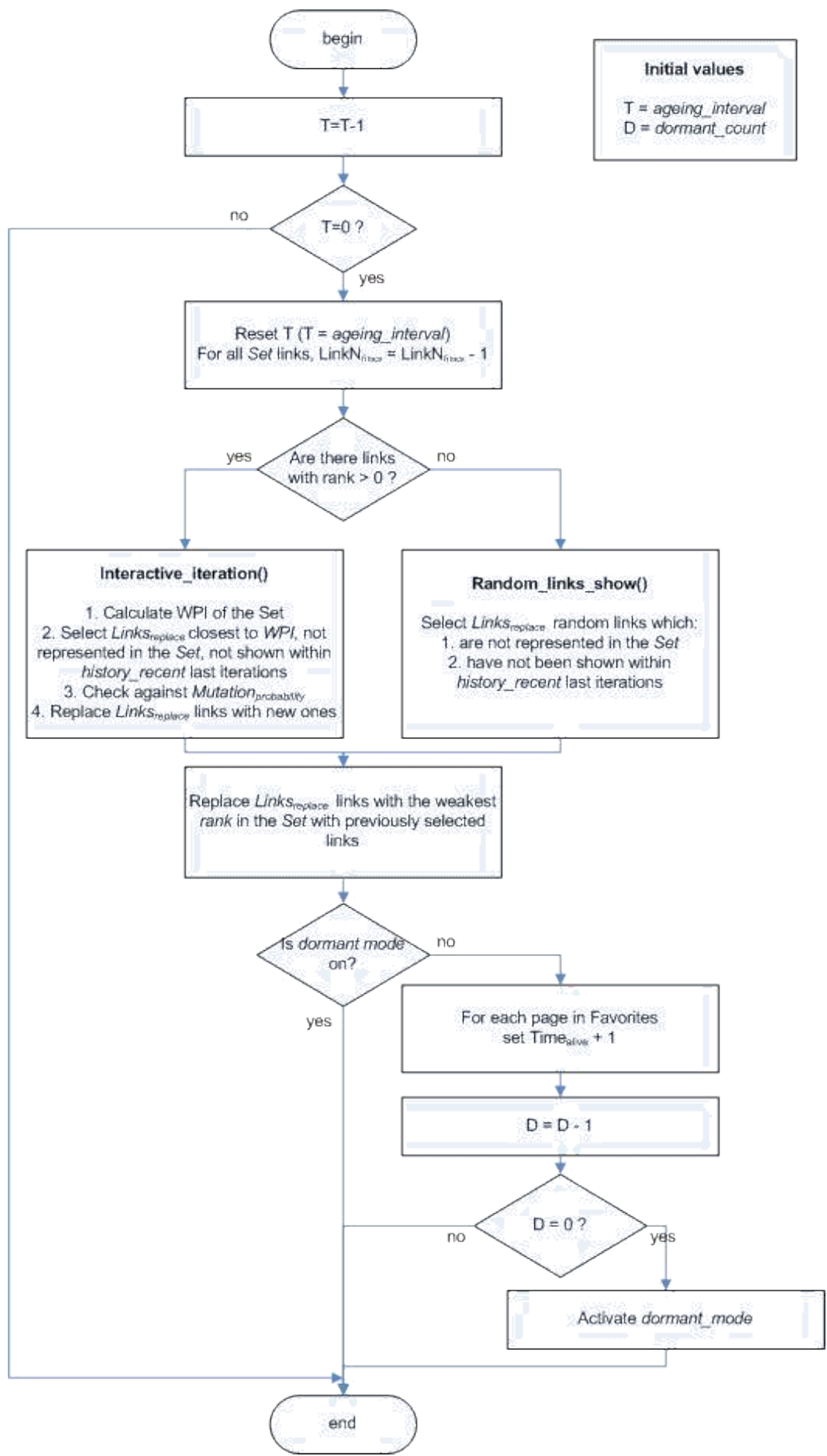

Figure 3.2. Periodic automatically launched algorithm.

The second flowchart represents algorithm of system decisions when user is taking some action:

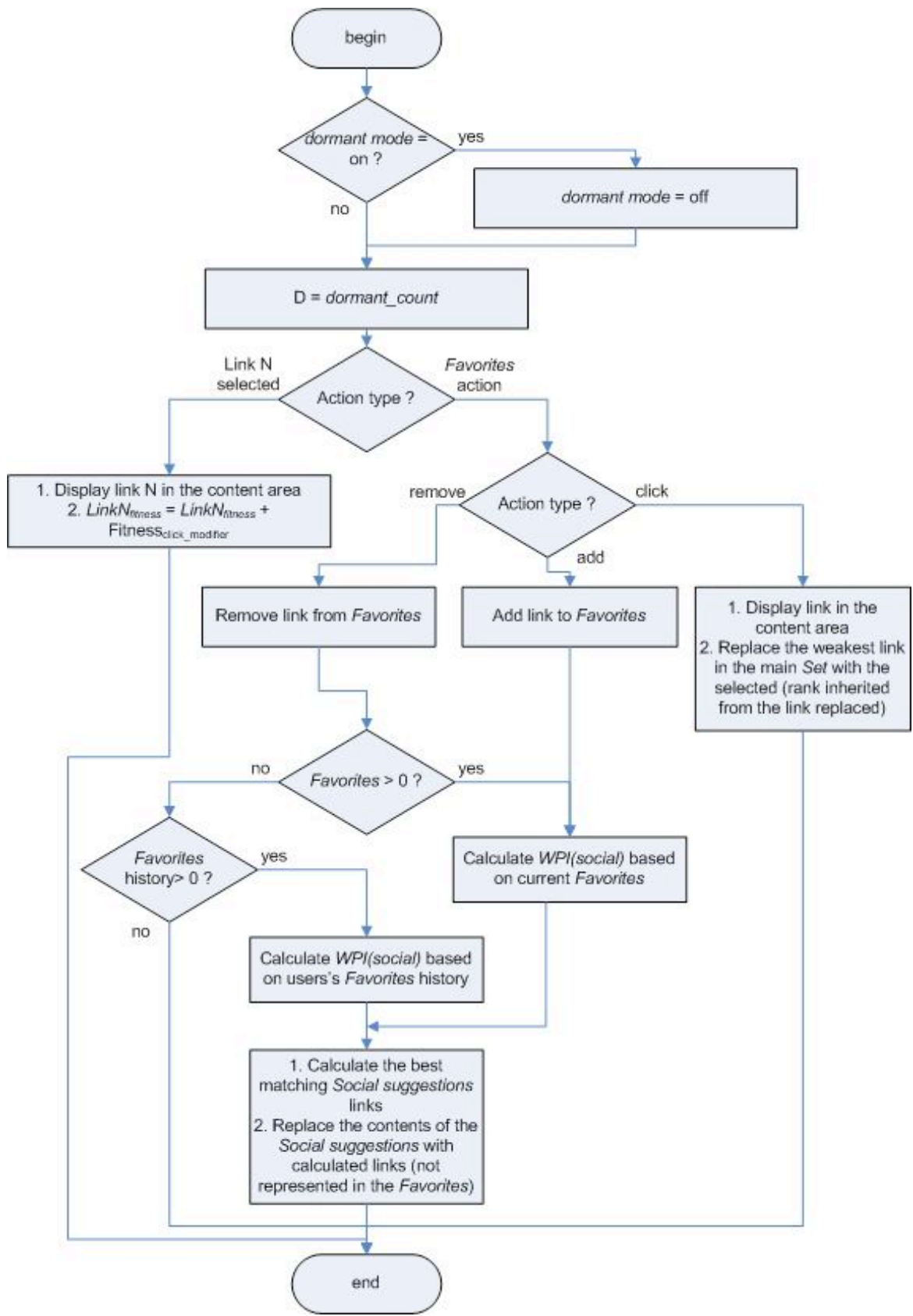

Figure 3.3. User's action algorithm.

## 4 Experiment

Current experiment has been conducted in order to evaluate our proposed data collection and processing architecure which involves converting documents corpus into vector space via tf.idf metrics methodology and then compressing the vector space representation with the help of dimensionality reduction techniques. The purpose of such processing was to:
1. Allow the system establish initial categorization of the corpus by achieving mathematically computable vector representations of all documents via tf.idf metric.
2. Allow the system to perform complex real-time calculations during each iteration comparing relevance between numerous documents in the corpus and taking into account. This has been achieved through significant minimization of feature set (vector size) with dimensionality reduction technique, PCA in current implementation.

For the purpose of evaluation of the approach, an extensive survey has been conducted where users have been continuously asked to evaluate the relevance between two random documents of the corpus and then the obtained relevance matrix has been compared to automated evaluations of the system with different parameters (no dimensionality reduction and various dimensionality reduction techniques with different parameters applied).

As a data source, the web site of Wessex Institute of Technology has been indexed with a limit of 3000 pages. The vocabulary of unique keywords, after stemming and stop words filtering obtained a size of 3897.

There have been conducted a survey in order to collect users' evaluation of relevance between 15 pre-selected pages in the corpus of documents used for evaluation in current work. 37 users with various level of knowledge in the area have left 352 opinions.

The correlation between relevancies reflected by Euclidian distances in SOM mappings and average pairwise relevancies obtained from survey results have been calculated. For comparison, in similar way correlation between SOM mappings and initial vector space distances have been also estimated. For comparison with other dimensionality reduction methods, Principal Component Analysis (PCA), Local Tangent Space Analysis (LTSA) and Stochastic Proximity Embedding (SPE) have been used. We list the results here in Table 3.2, techniques with different parameters given in the order of their performance:

Table 4.1: Various techniques compared with survey data

| Technique | Dimensionality | Correlation with survey data |
|---|---|---|
| PCA optimal | 19 | 0,445237453946 |
| Initial tf.idf vectors | 3897 | 0,42717044595324 |
| SPE 3d | 3 | 0,368396030076 |
| PCA 2d | 2 | 0,356945818598 |
| SPE 2d | 2 | 0,354836285591 |
| PCA 3d | 3 | 0,35202141743 |
| SOM 3d 10x10x10 | 3 | 0,309568117231 |
| SOM 3d 5x5x5 | 3 | 0,267381854644 |
| SOM 2d 40x40 | 2 | 0,254117116968 |
| LTSA 3d | 3 | 0,15592757837 |
| SOM 2d 20x20 | 2 | 0,128857727323 |
| SOM 3d 15x15x15 | 3 | 0,0744928015985 |
| LTSA 2d | 2 | -0,0116113046364 |

The results for discrete SOM measurements with various parameters for 2 and 3 dimensions in comparison to initial tf.idf and continuous (1-36 dimensions) measurements of PCA are graphically represented at Figure 4.1:

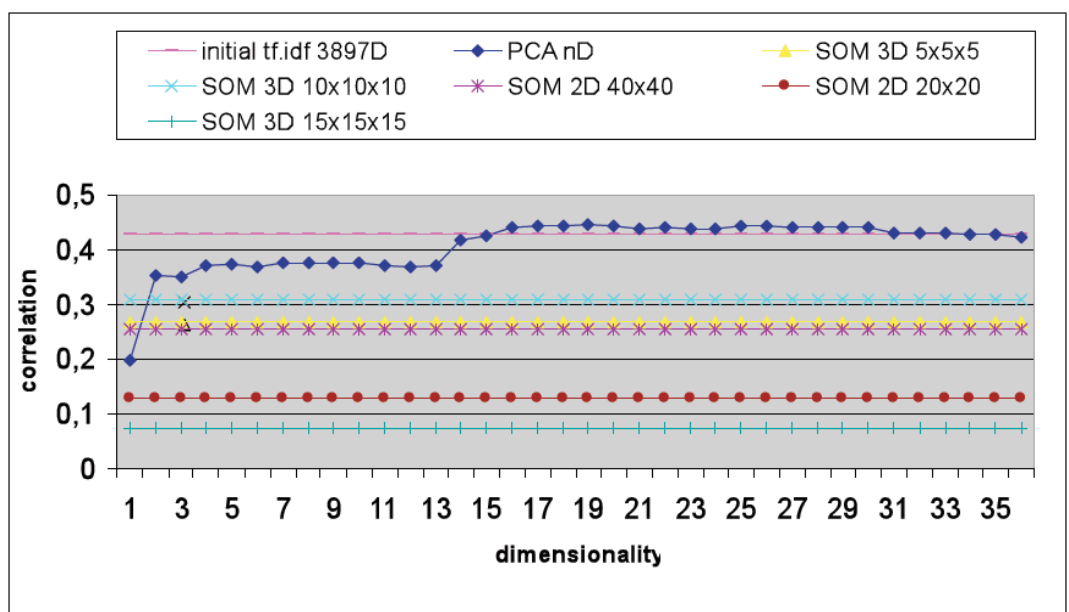

Figure 4.1: Mapping comparison chart

It can be seen from results that best performance on correlation is given by high-dimensional mappings: initial vectors (3897 dimensions) and PCA (best result in 19 dimensions). However from the point of view of visualisation such high-dimensional data is of no use. The results in 2d and 3d are quite comparative for all techniques.

It is obvious from the chart at Figure 4.1 that at 2-3 dimensional level a linear dimensionality reduction technique (PCA) does not outperform SOM significantly. For clarity at the chart given we use continuous lines for initial vectors and SOM despite their values correspond to dimensions 2 and 3 only. We can also see from results that both linear (PCA) and non-linear (SPE) are applicable and provide good results. The exception is LTSA giving poor results, which is to testify that tangent spaces analysis is not a successful approach in such a case.

It is interesting that optimal PCA outperforms even initial tf.idf data. It is out of scope of current work to establish whether this is fortuitousness or is it an evidence of the fact that during the process of mapping an optimal representational space have been found and the features have been automatically discovered which are optimal from the point of view of relevance calculation. In such case it would be possible to use dimensionality reduction as a complementary technique to tf.idf to extract most representative features of the corpus and establish unified mapping for all the documents. However the latter is only applicable on condition that the intrinsic dimensionality of the manifold in corpus data is determined which is presently a challenging task.

Comparing SOM with other techniques at dimensions 2 and 3 we can see that the performance is slightly lower than PCA and SPE still it is comparable and correlation remains at 'positive medium' level which demonstrates a strong dependency with commonsense estimation.

**Conclusions of the experiment**. By no means the described experiment should be considered as an ultimate comparison of the techniques. The goal of the experiment has not been to find the best universal dimensionality reduction technique for the information retrieval data in general but to study the behaviour of these techniques and different approaches in an existing real-world situation where the accuracy of knowledge representation provided by the implemented system significantly depends on the possibility of the compressed addressing space to keep the useful features of the original space and when such estimations given by the system could be compared to 'common sense' evaluations given by human users with average – above average level of knowledge of the field. It is important to understand that in other initial

conditions – different knowledge topic, data corpus, configuration of dimensionality reduction algorithms we could have obtained the very different results. As an example of how important the configuration is, the results of SOM techniques could be taken. The experiment has shown a strong influence of SOM configuration, in particular, dimensionality, on its performance. Thus, 3d SOM outperforms 2d SOM even when the number of neurons is lesser (40x40 performance still lower than 5x5x5). At the same time SOM 15x15x15 performs very poorly witnessing that the dependence is non-linear. These facts prove that configuration and architecture is very important for the performance of dimensionality reduction techniques implementations and researchers should experiment with different parameters (in case with SOM, such parameters are: dimensionality, number of neurons, topology, neighbourhood radius etc) in order to find the most effective configuration. Still the experiment allows us to conclude with the following general findings:

- The correlation of relevance calculated using uncompressed tf.idf method with users opinions data is medium positive (42%).
- The correlation of the idem space compressed through selected dimensionality reduction method (PCA) remains medium positive and estimates 35-44% therefore making such processing worthwhile in order to reduce calculations during real-time evaluations of relevance in the system.

There are other advantages of the implementation of the dimensionality reduction stage being useful in prospective:

- Out-of-sample selection is supported i.e. when a web page is added to the collection, the neural network saved in the database is able to determine a best location for a new coming document in the existing mapping space, there is no need to restart the mapping or reindex the pages.
- Upon such requirements it is possible to map initial data into either discreet (integer) or continuous space. In first case SOM technique should be used. In case continuous mapping space is required, standard techniques such as PCA should be used.

It is also interesting that an issue of intrinsic dimensionality is being broached by the experiment. The task of finding intrinsic dimensionality is still non-trivial for IR field. Otherwise it would have been possible to theorize regarding extraction of optimal relevance distinguishing features of the corpus. An evidence for that at particular dimensionality PCA outperforms even initial tf.idf data.

## 5 Conclusions

In the current work we explicitly describe the algorithms powering the system of collaborative study of web documents. The main advantages of the system proposed compared to modern search engines and knowledge base interfaces are the following:

1) Browsing approach. Using browsing rather than indexing approach we provide users with a more natural way of locating required documents. The user always has a fixed number of links to choose from and by clicking the most relevant ones he/she is able to reach the targeted documents. In such case it is not necessary for a user to know the title of the document or any key phrases as the browsing is being done following the contextual relevance chains. In many cases this approach is more beneficial than linguistic search through indexing as applied in modern search engines.

2) Intelligent evolutionary algorithm powered navigation. With the help of evolutionary algorithm it is possible to use a single navigational panel of a limited size to display links to all the documents in the corpus. It is an important advantage of the system that no manual pre-processing and categorization of documents corpus is required. System establishes initial relevance structure automatically and then refines it studying the documents access patterns of all users. The panel is dynamic and links to be shown are filtered according to latest real-time knowledge available to the system and previous interests expressed by current user.

3) Homogenous *Knowledge Map* space allowing simple mathematical calculations of the contextual relevance between documents. In our system each document obtains its coordinate in multidimensional space using the tdf.if metrics [26]. It is then possible to find out relevance by calculating the Euclidian distance between two certain documents. Moreover, it is possible to build complex requests 'find document Z which is relevant to X and 3 times more relevant to Y". Finally, the *Knowledge Map* concept allows easy mathematical representation of the current and previous interests of a certain user, which is called a *Weighted Point of Interest* (*WPI*) in our system. All the abovementioned parameters are used widely in the presented algorithms. To enable real time calculations we have proposed, implemented and evaluated through experiment the dimensionality reduction approach. The Homogenous Knowledge Map or Unified Knowledge Map concept can potentially introduce the next generation of global information indexation mechanism for mankind in replacement of current search engines using the "telegraph" keyword search approach. Once address space is unified an all human and computer agents can quickly address the same meaning via a unique address, a "knowledge link" (we could call it "k-link" or "'klink") which is simply a coordinates vector, this enables many useful outcomes that are going to boost productivity and further research and development in following areas: information retrieval, scientific research, projects collaboration (any type of projects), patent search and other information retrieval tasks for large unstructured datasets, information classification and exchange in any human work areas, but also for the purposes of AI (artificial intelligence) as this allows computer agents to automatically assign meaning to any knowledge and address documents and information pieces in a unified way, quantifying their meaning and being able to mathematically calculate the relevance between ANY two documents in the world. We believe this is going to further boost the development of artificial intelligence and autonomous computer agents therefore. Additionally, thanks to a unified and universal global address space, there is a huge potential in the development of brain-computer interfaces and 'digital telepathy'. With the help of fMRI imaging, human thoughts or formulated intentions could be translated with the help of e.g., artificial neural network classifier, and then the global mapping in our proposed knowledge map is going to be established as a vector space numerical address. It is then a straightforward task to transfer the same numerical code to a computer or other human (equipped potentially with a fMRI transmitter or a visual interpreter such as computer display or Google Glass type interface). This way meanings can be transferred back and forth between humans and computers and humans and humans without the need to use keyboard or mouse for inputting information. This allows for a precise transfer of meanings once and for all eliminating the time and, most importantly, context losses when using "telegraph" keyword based search engines.

    4) Social suggestions. The history of confirmed interests (*Favorites* mechanism) is being stored for each user. This and other individual parameters are normalized during calculations. The *dormant mode* feature tracks the periods of inactivity. The abovementioned *WPI* method allows real time calculation of current user's interests and those of other users. In combination these allow finding the documents which should be of most interest for the current user, based on data mining performed automatically by other users while interacting with the system.

    The experimental implementation of the system has proven the applicability of the proposed combination of algorithms and methods. The results of users survey display good correlation of automated estimations with human common-sense estimations.

    The main aim of our work was to propose a systematized method to be used in the industry of search, information retrieval and knowledge representation. Further research and improvements as well as practical applications are encouraged. The possible fields of application vary widely, from traditional web search where the system could be used to refine results to topic oriented knowledge bases for communities of experts or self organized web portals. Due to its browsing approach, high level of user-adaptability and some innovative features such as *Knowledge Map*, the system might find successful applications in many fields linked with data processing, either on its own or in combination with existing systems and methods.

    Further work possibilities are broad. The necessity in certain improvements and modifications may vary depending on current implementation and application field. These are the major points in our method, which could be improved, worked on or modified depending on

application, as from our point of view:
- Evaluate different implementations and variations of td.idf metrics for *Knowledge Map* generation;
- Consider alternative (to tf.idf metrics) methods for *Knowledge Map* generation. Evaluate application of text recognition, ontology models and other alternative approaches (on their own or in combination);
- Further evaluate different dimensionality reduction methods for the *Knowledge Map* space, implement automated intrinsic dimensionality calculation, study the effects in relevance calculation improvements;
- Evaluate different clustering methods for documents coordinates in the *Knowledge Map* space;
- Evaluate the option of introducing clustering for users into groups of interests.

The other area for improvements in the implementation is interface. Compared to our experimental implementation we expect the versions applied to real world problems to have multiple improvements in terms of interface design and usability as well of code optimization making it more convenient for the users to use the system and easier for the server to handle substantial loads.

## 5 References

[1] F. Abbatista, A. Paradiso, G. Semerano, F. Zambetta, An agent that learns to support users of a Web site. Applied Soft Computing, 4 (2004) 1-12.

[2] J. Allan et al, Challenges in information retrieval and language modeling: report of a workshop held at the center for intelligent information retrieval, University of Massachusetts Amherst, September 2002, ACM SIGIR Forum 37 (1) (2003) 31-47.

[3] G.A. Alvarez, S.L. Franconeri,. How many objects can you track?: Evidence for a resource-limited attentive tracking mechanism, Journal of Vision, 7(13):14 (2007) 1-10.
Retrieved August 2008, from http://www.journalofvision.org/7/13/14/

[4] A.N. Badre, Shaping Web Usability: Interaction Design in Context, Addison Wesley Professional, Boston, MA, 2002.

[5] M. Bernard, Examining User Expectations for the Location of Common E-Commerce Web Objects, Usability News, 4(1) (2002). Retrieved July 2008, from http://www.surl.org/usabilitynews/41/web_object-ecom.asp

[6] M. Bernard, S. Hull and D. Drake, Where should you put the links? A Comparison of Four Locations. Usability News, 3 (2) (2001).
Retrieved July 2008, from http://www.surl.org/usabilitynews/32/links.asp

[7] M. Bernard, L. Larsen, What is the best layout for multiple-column Web pages? Usability News, 3 (2) (2001)
Retrieved July 2008, from http://www.surl.org/usabilitynews/32/layout.asp

[8] M. D. Byrne, J.R. Anderson, S. Douglass and M. Matessa (1999). Eye tracking the visual search of click-down menus, In: Proc. CHI'99, pp. 402-409.

[9] M. Chau, D. Zeng, H. Chen, M. Huang, D. Hendriawan, Design and evaluation of a multi-agent collaborative Web mining system, Decision Support Systems: Web retrieval and mining, 35 (1) (2003), 167-183.

[10] S. Cheng, Y. Wang, Z. Wu, Provable Dimension Detection using Principal Component Analysis, In: Proc. 21$^{st}$ annual symposium on Computational geometry, Pisa, Italy, 2005, pp. 208-217.

[11] B.D. Ehret, Learning where to look: Location learning in graphical user interfaces, In: Conf. Proc. CHI 2002, pp. 211-218.

[12] D.K. Farkas, J.B. Farkas, Guidelines for designing web navigation. Technical Communication, 47(3) (2000) 341-358.

[13] A.J. Hornof, T. Halverson, Cognitive strategies and eye movements for searching hierarchical computer displays, In: Conf. Proc CHI 2003, pp. 249-256.

[14] J. Kalbach, T. Bosenick, Web page layout: A comparison between left and right-justified site navigation menus, Journal of Digital Information, 4(1) (2003).
Retrieved August 2008, from http://jodi.tamu.edu/Articles/v04/i01/Kalbach/.

[15] M. Khordad, M. Shamsfard & F. Kazemeyni, A hybrid method to categorize HTML document, In: Proc 6$^{th}$ Int. Conf. Data Mining, Text Mining and their Business Applications, 2005, pp. 331-340.

[16] J.R. Kingsburg, A.D. Andre, A comparison of three-level web menus: Navigation structures, In: Proc. The Human Factors and Ergonomics Society Annual Meeting (2004)

[17] S. Kumar, V. S. Jacob and C. Sriskandarajah, Scheduling advertisements on a web page to maximize revenue, European journal of operational research, 173 (2006) 1067-1089.

[18] L.J.P. van der Maaten, An Introduction to Dimensionality Reduction Using Matlab, Technical Report MICC-IKAT 07-06, Maastricht University, Maastricht, The Netherlands, 2007. Retrieved August 2008 from http://www.cs.unimaas.nl/l.vandermaaten/Laurens_van_der_Maaten/Publications_files/Demonstration1.pdf

[19] C. T. Meadow, Text information retrieval systems, Academic Press, San Diego, CA, 1992

[20] F. Menczer, G. Pant, P. Srinivasan, Topical web crawlers: Evaluating adaptive algorithm, ACM Transactions on Internet Technology (TOIT), 4 (4) (2004) 378-419.

[21] M. Niemela, J. Saarinen, Visual search for grouped versus ungrouped icons in a computer interface. *Human Factors*, 42(4) (2000) 630-635.

[22] V.L. O'Day, R. Jeffries, Information artisans: patterns of result sharing by information searchers, In: Proc. of the ACM Conf. on Organizational Computing Systems, COOCS, Milpitas, CA, 1993, pp. 98– 107.

[23] C.J. van Rijsbergen, S.E. Robertson and M.F. Porter, New models in probabilistic information retrieval, British Library (British Library Research and Development Report, no. 5587), London, 1980

[24] M. Sahami, S. Yusufali, M. Baldonado, SONIA: a service for organising networked information autonomously, In: Proc. 3$^{rd}$ ACM Conf. Digital libraries, Pittsburgh, Pennsylvania, United States, 1998, pp. 200-209.

[25] G. Salton, C. Buckley, Term-weighting approaches in automatic text retrieval, Information Processing & Management, 24 (5) (1998) 513–523.

[26] F. Sebastiani, Machine learning in automated text categorization, ACM Computing Surveys (CSUR), 34 (1) (2002) 1-47.

[27] N. Stojanovic, A. Maedche, S. Staab, R. Studer, Y. Sure, SEAL – A Framework for Developing SEmantic PortALs, In: Proc. 1$^{st}$ Int. Conf. Knowledge capture, 2001, pp. 155-162.

[28] T.Yan, M.Jacobsen, H. Garcia-Molina, U.Dayal, From user access patterns to dynamic hypertext linking, In: Proc. 5$^{th}$ Int. WWW Conf. Computer Networks and ISDN, 1996, pp. 1007-1014.